\documentclass[a4paper,10pt,twoside]{cpc-hepnp}

\usepackage{multicol}
\usepackage{graphicx}
\usepackage{booktabs}
\usepackage{amssymb,bm,mathrsfs,bbm,amscd}
\usepackage[tbtags]{amsmath}
\usepackage{lastpage}
\usepackage{CJK}
\usepackage{color}
\usepackage[colorlinks=true, linkcolor=red, citecolor=blue,CJKbookmarks=true]{hyperref}
\usepackage{times}

\graphicspath{{figs/}}

\def\TeV{\mathrm{TeV}} % TeV
\def\GeV{\mathrm{GeV}} % GeV
\begin{document}
%\begin{CJK*}{GBK}{song}

\fancyhead[c]{\small Chinese Physics C~~~Vol. xx, No. x (2020)
085001} \fancyfoot[C]{\small 085001-\thepage}

\footnotetext[0]{Received 22 October 2019}

\title{Expected LHAASO sensitivity to decaying dark matter signatures from dwarf galaxies gamma-ray emission\thanks{Supported by the National Key R~$\&$~D Program of China (Grant No. 2016YFA0400200), the National Natural Science Foundation of China (Grant Nos.
U1738209, 11851303, 11835009, 11975072), and the National Program for Support of Top-Notch Young Professionals. }}

\author{
Dong-Ze He$^1$
\quad
Xiao-Jun Bi$^{2,3}$
\quad
Su-Jie Lin$^{2;1)}$\email{linsj@ihep.ac.cn}
\quad
Peng-Fei Yin$^2$
\quad
Xin Zhang$^{1,4;2)}$\email{zhangxin@mail.neu.edu.cn}
}

\maketitle

\address{%
{$^1$Department of Physics, College of Sciences, Northeastern University, Shenyang 110819, China}\\
{$^2$Key Laboratory of Particle Astrophysics, Institute of High Energy Physics, Chinese Academy of Sciences, Beijing 100049, China}\\
{$^3$School of Physical Sciences, University of Chinese Academy of Sciences, Beijing 100049, China}\\
{$^4$Center for High Energy Physics, Peking University, Beijing 100080, China}
}

\begin{abstract}
As a next-generation complex extensive air shower array with a large field of view, the large high altitude air shower observatory (LHAASO) is very sensitive to the very high energy gamma-rays from $\sim$ 300 GeV to 1 PeV, and may thus serve as an important probe for the heavy dark matter (DM) particles.
In this study, we make a forecast for the LHAASO sensitivities to the gamma-ray signatures resulting from DM decay in dwarf spheroidal satellite galaxies (dSphs) within the LHAASO field of view.
Both individual and combined limits for 19 dSphs incorporating the uncertainties of the DM density profile are explored.
Owing to the large effective area and strong capability of the photon-proton discrimination, we find that LHASSSO is sensitive to the signatures from decaying DM particles above $\mathcal{O}(1)$ TeV.
The LHAASO sensitivity to the DM decay lifetime reaches $\mathcal{O} (10^{26}) \sim \mathcal{O} (10^{28})$ s for several decay channels at the DM mass scale from 1 TeV to 100 TeV.
\end{abstract}


%\begin{pacs}

%\end{pacs}

\footnotetext[0]{\hspace*{-3mm}\raisebox{0.3ex}{$\scriptstyle\copyright$}
2020 Chinese Physical Society and the Institute of High Energy Physics of the Chinese Academy of Sciences and the Institute of Modern Physics of the Chinese Academy of Sciences and IOP Publishing Ltd
}%

\begin{multicols}{2}

\section{Introduction}
The cosmological constant $\Lambda$ and cold dark matter (DM) paradigm have made numerous far-reaching predictions about the composition of the Universe.
An abundance of compelling observational evidence has been accumulated to account for the presence of DM.
DM should be neutral, non-baryonic, and cold, and constitute nearly 84\% of the total matter of the universe \cite{Adam:2015rua}.
However, little is known about the DM microscopic properties as an elementary particle.
To understand the particle nature of DM, numerous new physics models have been proposed in the literature \cite{Bertone:2004pz}, among which the weakly interacting massive particles (WIMPs) approach is the most attractive candidate.

%Candidates of WIMPs are predicted in the extensions of the Standard Model (SM) of particle physics, for instance the lightest supersymmetric
%particle (neutralino) from the theory of supersymmetry \cite{Jungman:1995df} and the lightest Kaluza$-$Klein particle from the theory of universal extra dimensions \cite{Servant:2002aq}. Both of these hypothetical particles could interact with SM particles via a weak-scale force.
WIMPs could either decay or self-annihilate into steady standard model (SM) particles through some weak interactions, such as gamma-rays, neutrinos, and anti-matter particles.
Indirect DM detection is performed by experiment that investigate such high energy signals.
In particular, the gamma-ray signal is a powerful probe to reveal the properties of DM owing to its simple propagation process.
Among the astrophysical sources with high DM densities, dwarf spheroidal galaxies (dSphs) are the most promising research objects in the search for gamma rays emitted from DM \cite{Albert:2017vtb,Baring:2015sza,Aliu:2012ga,Halder:2019pro,Ackermann:2015zua,Morselli:2017ree,Acharya:2017ttl}.
These sources are relatively nearby, highly DM dominated with a large order of magnitude of the mass-to-light ratio $\mathcal{O} (10-100)$, and almost free of astrophysical backgrounds~\cite{Mateo:1998wg,Grcevich:2009gt}.
With these outstanding advantages, dSphs would offer the cleanest DM signals compared with other objects.

Currently, gamma ray astronomy above tens of TeV remains almost completely unexplored, as past and present telescopes can only record few photons in this energy range.
A strong interest in the very high energy (VHE) gamma-ray astronomy was aroused by the development of next-generation instruments, which are capable of more sensitive observations with a larger field of view (FOV) in a more extended energy region.
This interest brought on the ambitious project of the large high altitude air shower observatory (LHAASO) still under construction, which aims to cover the energy range approximately from 300 GeV to 1 PeV \cite{Bai:2019khm}.
Remarkably, the design concept of LHAASO is to make this continuously-operated instrument extremely competitive for the gamma-ray observation in the energy range above tens of TeV.
Therefore, through the VHE gamma-ray observation from dSphs by LHAASO, it is compelling to search for the DM signatures and set strong limits on the properties of heavy DM particles.

In Ref. \cite{He:2019dya}, we investigated the expected sensitivities of the LHAASO project to gamma-ray signals induced by self-annihilating DM particles in 19 selected dSphs.
Although it is natural to assume that the DM particles are absolutely stable, this assumption is not necessary.
In fact, the current cosmological and astrophysical observations only require that the lifetime of DM particles is significant longer than the age of the Universe, about 13.8 Gyr ($\rm 4.56\times 10^{17}\,s$).
This long lifetime can be achieved by some interactions at high energy scales; the relevant signatures may be detectable by indirect detection experiments (see, e.g. Refs. \cite{Yin:2008bs,Ibarra:2008jk,Nardi:2008ix,Arvanitaki:2008hq,Ibarra:2013cra} and references therein).
Thus, the LHAASO gamma-ray observation of dSphs can also search for the signals originating from DM decay.

In this study, as a further step along this line, we perform a forecast of the LHAASO sensitivity to the lifetime of decaying DM using the mimic observation of VHE gamma-rays for 19 dSphs within the LHAASO FOV.
In the analysis, we take the statistic uncertainties of the spatial DM distribution of dSphs into account \cite{Geringer-Sameth:2014yza,Pace:2018tin,Hayashi:2016kcy}.
To derive a reasonable sensitivity, the simulated data of LHAASO, considering its strong background rejection power, are utilized.

This paper is organized as follows.
In Sec. \ref{sec:gamma-flux}, we introduce the calculation of the gamma-ray flux from DM decay.
In Sec. \ref{sec:results}, we show the LHAASO sensitivities and provide comparisons with other experimental results.
Finally, the conclusion is presented in Sec. \ref{sec:conclu}.

\section{Gamma-ray signals from DM decay in dSphs}\label{sec:gamma-flux}
In this study, we assume dSphs to be point-like sources.
The expected gamma-ray flux resulting from DM decay in a point-like source is described by
\begin{equation}\label{func:decay-flux}
\Phi=\frac{1}{4\pi}\frac{1}{m_{\chi} \tau}\int^{E_{\rm max}}_{E_{\rm min}}\frac{dN_{\gamma}}{dE_{\gamma}}dE_{\gamma} \times D,
\end{equation}
where $m_{\chi}$ is the mass of DM particles; $\tau$ is the decay lifetime of DM particles; the integration is performed over each energy bin between $E_{\rm min}$ and $E_{\rm max}$; and $\frac{dN_{\gamma}}{dE_{\gamma}}$ denotes the gamma-ray differential energy spectrum resulting from the decay of a DM particle via a certain final state channel.
In this study, we derive $\frac{dN_{\gamma}}{dE_{\gamma}}$ with the utilization of the {PPPC4DM} package \cite{Cirelli:2010xx,Ciafaloni:2010ti}.

In Eq.~(\ref{func:decay-flux}), the astrophysical factor ``$D$ factor" is an integral of the DM density along the line of sight ($\rm l.o.s$) distance $x$ in the region of interest
\begin{equation}\label{func:dfactor}
D=\int_{\rm source}d\Omega\int_{\rm l.o.s} dx\rho(r(\theta,x)),
\end{equation}
where the solid angle $\Omega$ varies in the observed regions with an integration angle $\Delta \Omega=2\pi\times[1-\cos\alpha_{\rm int}]$, and $\rho{(r)}$ describes the DM density profile of the astrophysical system varying with the distance $r$ from its center.
The DM density profile of dSphs can be determined by the Jeans equation using the kinematic observation of stellar velocities (see e.g., Refs.
\cite{Evans:2003sc,Strigari:2007at,Martinez:2009jh}).

In this analysis, we also consider the statistical uncertainty of the $D$ factor of the dSph employing the method of Refs.~\cite{Ackermann:2015zua,Fermi-LAT:2016uux}.
The likelihood in all energy bins for one dSph is given by
\begin{equation}
\mathcal{L}_{j}=\prod_{i}\mathcal{L}_{ij}(S_{ij}|B_{ij},N_{ij})\times\dfrac{e^{-[\rm log_{10}(D_{\it j})-log_{10}(D_{obs,{\it j}})]^{2}/2\sigma_{\it j}^{2}}}{{\rm ln}(10)D_{{\rm obs},j}\sqrt{2\pi}\sigma_{j}}.
\label{likelihood}
\end{equation}
{Here, $\mathcal{L}_{ij}$ is the likelihood that is taken to be the Poisson distribution,
\begin{equation}\label{func:likelihood-l}
  \mathcal{L}_{ij}(S_{ij}|B_{ij},N_{ij})=\prod_{i}\frac{(B_{ij}+S_{ij})^{N_{ij}} {\rm{exp}} [-(B_{ij}+S_{ij})]}{N_{ij}!},
\end{equation}
where $S_{ij}$, $B_{ij}$, and $N_{ij}$ denote the numbers of the expected signal counts from the DM decay, expected background counts from cosmic rays, and total observed counts in the $i$-th energy bin for the $j$-th dSph, respectively. Because the value of $S_{ij}$ is physically restricted to be equal or greater than zero, for energy bins with observed counts under the statistic fluctuations of the background, the value of $S_{ij}$ maximizing the likelihood is supposed to be zero. This is consistent with the fact that no gamma photons detected from the DM sources.}
Furthermore, $\rm log_{10}(D_{obs,{\it j}})$ and $\sigma_{j}$ denote the mean value and corresponding standard deviation of the $D$ factor, respectively.
For given $\tau$ and $m_{\chi}$ values, $\rm log_{10}(D_{\it j})$ is assumed to be the value maximizing the likelihood $\mathcal{L}_{j}$.
We take the calculated mean values of the $D$ factor and their statistical uncertainties of 19 dSphs from Refs.~\cite{Geringer-Sameth:2014yza,Pace:2018tin,Hayashi:2016kcy} and list them in Table \ref{Table:dsphs} .

\begin{table*}
\caption{\label{Table:dsphs}Astrophysical properties of 19 selected dSphs within LHAASO FOV.
Columns denote the name, right ascension (RA.), declination (DEC.), distance, effective time ratio ($r_{\rm eff}$), maximum angular angle $\rm \theta_{max}$, and $D$ factor for each dSph.
The $D$ factor and $\rm \theta_{max}$ of the dSphs are provided by Ref. \cite{Geringer-Sameth:2014yza},
except for the four dSphs marked with asterisks, for which the $D$ factors are not provided in this reference.
We adopt the $D$ factors from Ref. \cite{Pace:2018tin} for Draco II, Pisces II, Willman 1 and from Ref. \cite{Hayashi:2016kcy} for Triangulum II.}
\footnotesize\centering
\begin{tabular*}{170mm}{@{\extracolsep{\fill}}ccccccccccccccc} \toprule%\hline \hline
              &&
          RA. &&
          DEC.&&
          Distance&&
          $r_{\rm eff}$ &&
          $\theta_{\rm max}$&&
          $\log_{10}D_{\rm obs}$&&\\
		
       Source&&
		(deg)&&
		(deg)&&
        (kpc)&&
		  &&
		(deg)&&
		({$\rm log_{10}[\GeV\rm cm^{-2}$]})\\\hline

		$\rm Bo\ddot{o}tes ~I$&&
		$210.02$&&
		$14.50$&&
		66&&
        0.352&&
		$0.47$&&
		$17.9\pm0.2$\\

		Canes Venatici $\rm I$&&
		$202.02$&&
		$33.56$&&
		218&&
		0.398&&
        $0.53$&&
		$17.6\pm0.5$\\
		
		Canes Venatici~$\rm II$&&
		$194.29$&&
		$34.32$&&
		160&&
        0.399&&
		$0.13$&&
		$17.0\pm0.2$\\
		
		Coma~Berenices&&
		$186.74$&&
		$23.90$&&
		44&&
        0.377&&
		$0.31$&&
		$18.0\pm0.2$\\
		
		Draco&&
		$260.05$&&
		$57.92$&&
		76&&
        0.442&&
		$1.30$&&
		$18.5\pm0.1$\\

		Draco II$^{\star}$&&
		$238.20$&&
		$64.56$&&
		24&&
        0.451&&
		$-$&&
		$18.0\pm0.9$\\
		
        Hercules&&
		$247.76$&&
		$12.79$&&
		132&&
        0.348&&
		$0.28$&&
		$16.7\pm0.4$\\

        Leo I&&
		$152.12$&&
		$12.30$&&
		254&&
        0.346&&
		$0.45$&&
		$17.9\pm0.2$\\

        Leo II&&
		$168.37$&&
		$22.15$&&
		233&&
        0.372&&
		$0.23$&&
		$17.2\pm0.4$\\

        Leo IV&&
		$173.23$&&
		$-0.54$&&
		154&&
        0.303&&
		$0.16$&&
		$16.1\pm0.9$\\

        Leo V&&
		$172.79$&&
		$2.22$&&
		178&&
        0.314&&
		$0.07$&&
		$15.9\pm0.5$\\

        Pisces II$^{\star}$&&
		$344.63$&&
		$5.95$&&
		182&&
        0.327&&
		$-$&&
		$17.0\pm0.6$\\

        Segue 1&&
		$151.77$&&
		$16.08$&&
		23&&
        0.357&&
		$0.35$&&
		$18.0\pm0.3$\\

        Sextans&&
		$153.26$&&
		$-1.61$&&
		86&&
        0.299&&
		$1.70$&&
		$17.9\pm0.2$\\

        Triangulum II$^{\star}$&&
		$33.32$&&
		$36.18$&&
		30&&
        0.403&&
		$-$&&
		$18.4\pm0.8$\\

        Ursa Major I&&
		$158.71$&&
		$51.92$&&
		97&&
        0.432&&
		$0.43$&&
		$17.6\pm0.3$\\

        Ursa Major II&&
		$132.87$&&
		$63.13$&&
		32&&
        0.449&&
		$0.53$&&
		$18.4\pm0.3$\\

        Ursa Minor&&
		$227.28$&&
		$67.23$&&
		76&&
        0.455&&
		$1.37$&&
		$18.0\pm0.1$\\

        Willman 1$^{\star}$&&
		$162.34$&&
		$51.05$&&
		38&&
        0.430&&
		$-$&&
		$18.5\pm0.6$\\
    \bottomrule
    %\hline\hline
	\end{tabular*}
    \end{table*}

In the literature, two sets of $D$ factors are provided, depending on the choice of the integration angle.
One set is calculated within a constant integration angle, e.g. $\alpha_{\rm int}=0.5^\circ$.
The other set is derived within the maximum angular radius of the source ${\rm arcsin}(r_{\rm max}/d)$, where $r_{\rm max}$ is an estimate of distance from the dSph center to the outermost member star, and $d$ is the distance from the Earth to the source.
In general, the DM particles tend to contribute signals from the vicinity of the source center due to the density profile, while the angle distribution of the background resulting from cosmic rays is almost flat.
Therefore, to suppress the background, we adopt the $D$ factors integrated over a smaller angle region with $\alpha_{\rm int}=\min \{ \theta_{\rm max}, 0.5^\circ \}$.

\section{LHAASO Sensitivities to DM Lifetime}\label{sec:results}
The LHAASO experiment is being built on the HaiZi Mountain (4410 m a.s.l.) near Daocheng in the Sichuan province of China.
It comprises a complex extensive air shower (EAS) array consisting of three sub-arrays: the square kilometer particle detector array (KM2A), the water cherenkov detector array (WCDA), and the wide field Cherenkov telescope array (WFCTA).
WCDA and KM2A are designed for the photons with energies approximately from 100 GeV to 20 TeV and above 20 TeV, respectively, which are relevant for the gamma-ray detection.
Further details on the LHAASO experiment are provided in Refs.
\cite{Cao:2010zz,Cao:2014rla}.

When VHE gamma-rays and cosmic ray nuclei enter the atmosphere, they interact with the atmospheric nuclei and then separately generate the electromagnetic and hadron cascades, collectively known as EAS.
Subsequently, these secondary particles from EAS would impinge on the water Cherenkov detectors (WCDs) of LHAASO and produce Cherenkov lights.
The hadronic backgrounds and photon signals can be distinguished in light of their different energy distributions deposited across the WCDs.
Furthermore, for gamma rays above 10 TeV, the measurement of the muon component in the shower by the muon detectors of KM2A further allows a more efficient hadronic background rejection.

%In this section, we shall talk about the LHAASO sensitivity to the DM decay lifetime through the gamma-ray observation towards dSphs. The LHAASO simulated integral sensitivity curve to a Crab-like source is shown in FIG. \ref{fig:sensi}. It is the combination of two components, with the first relative to the WCDA, operating in the energy range from $300\,\GeV$ to $10\,\TeV$, and the second relative to the KM2A array, sensitive to the energy above 10 TeV. Moreover, for comparison, the sensitivity curves of other experiments are also exhibited in the same figure. We can clearly find, at the energy range above $\sim 10$ TeV, LHAASO is more sensitive than other ground-based experiments, which further manifests that the LHAASO project will be able to remain a more powerful capability to explore the property of heavy DM particle in the near future.%
We investigate the LHAASO sensitivity to the DM decay signals from 19 selected dSphs with large $D$ factors listed in Table \ref{Table:dsphs}.
In comparison with the research of HAWC \cite{Albert:2017vtb}, we adopt four more dSphs in the analysis, taking into account the larger FOV of LHAASO (defined in the declination range $-11^{\circ}<\delta<69^{\circ}$), including {Draco II, Leo V, Pisces II}, and {Willman 1}.
For each dSph, we perform a series of mimic observations under the null hypothesis and then calculate the likelihood described by Eq.~\ref{likelihood}.
Further details of the analysis are provided in Appendix.~\ref{section:method}, which are similar to our previous study \cite{He:2019dya}.
Subsequently, we derive the 95\% sensitivity to the DM lifetime $\tau$ by decreasing the likelihood by 2.71/2 from its maximum with the given DM mass in each mimic observation, assuming a $\chi^2$-distributed test statistic \cite{Rolke:2004mj}.

\begin{figure*}[!htbp]
\begin{center}
{\includegraphics[width=0.45\textwidth]{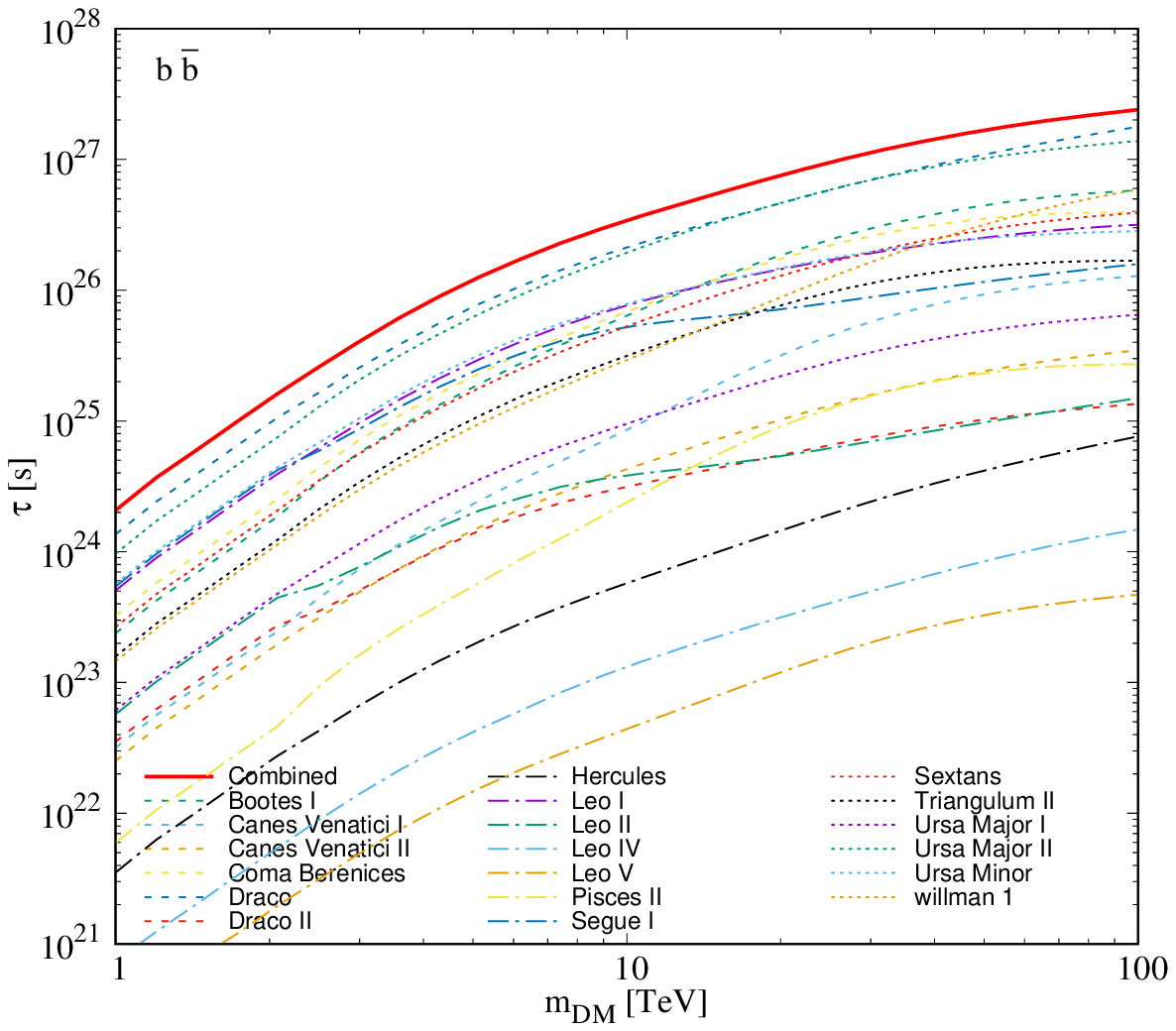}}
{\includegraphics[width=0.45\textwidth]{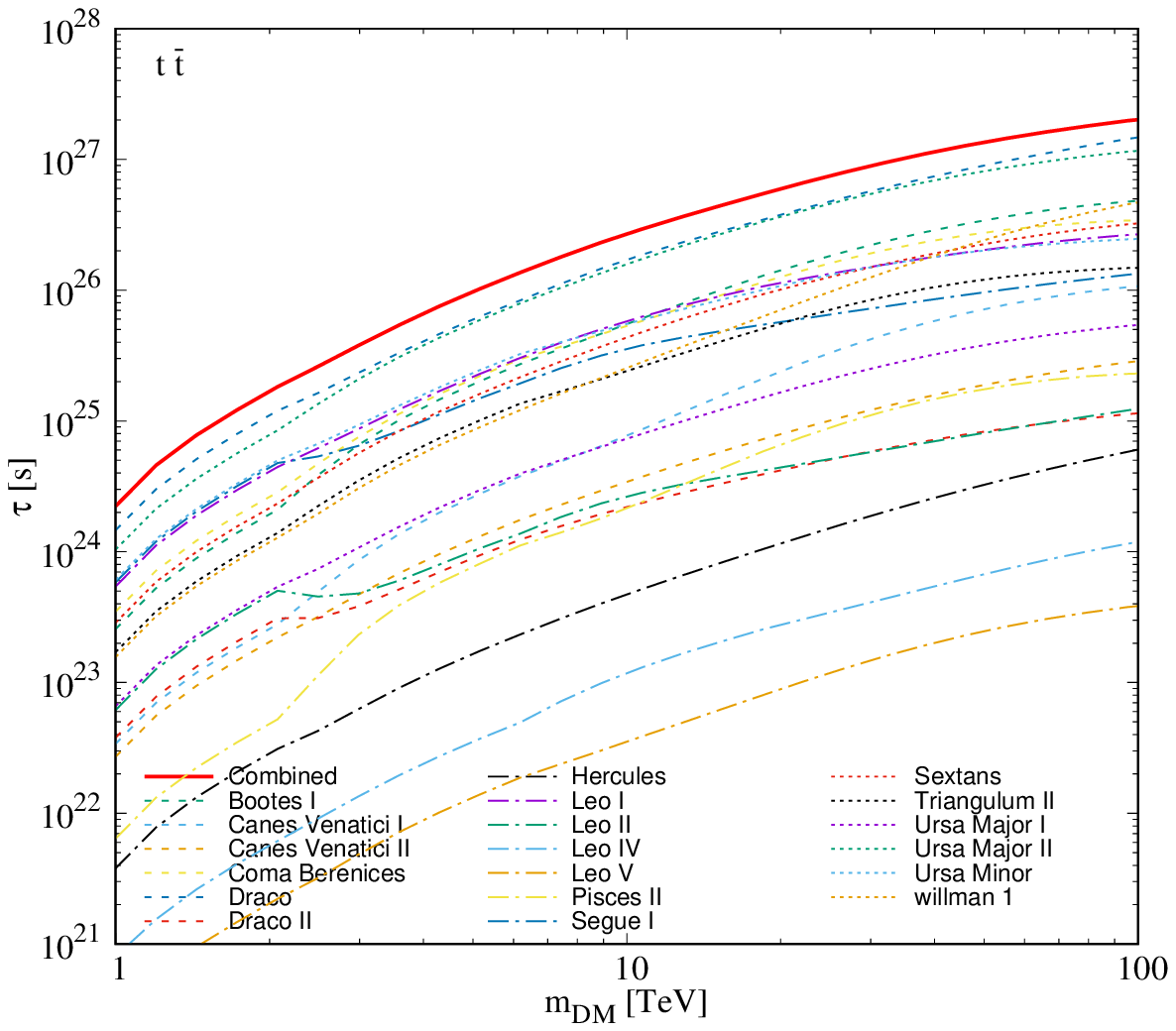}}
{\includegraphics[width=0.45\textwidth]{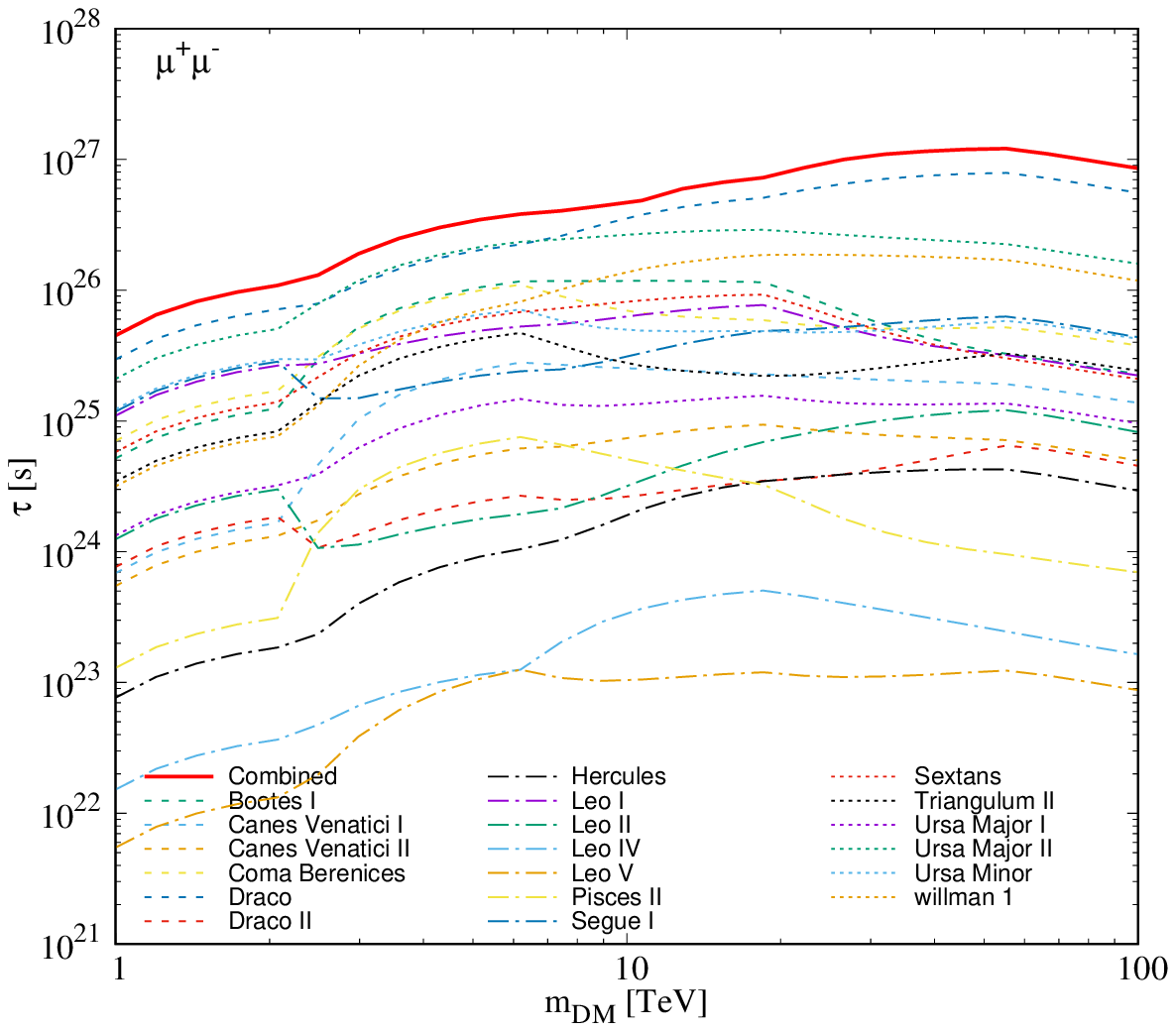}}
{\includegraphics[width=0.45\textwidth]{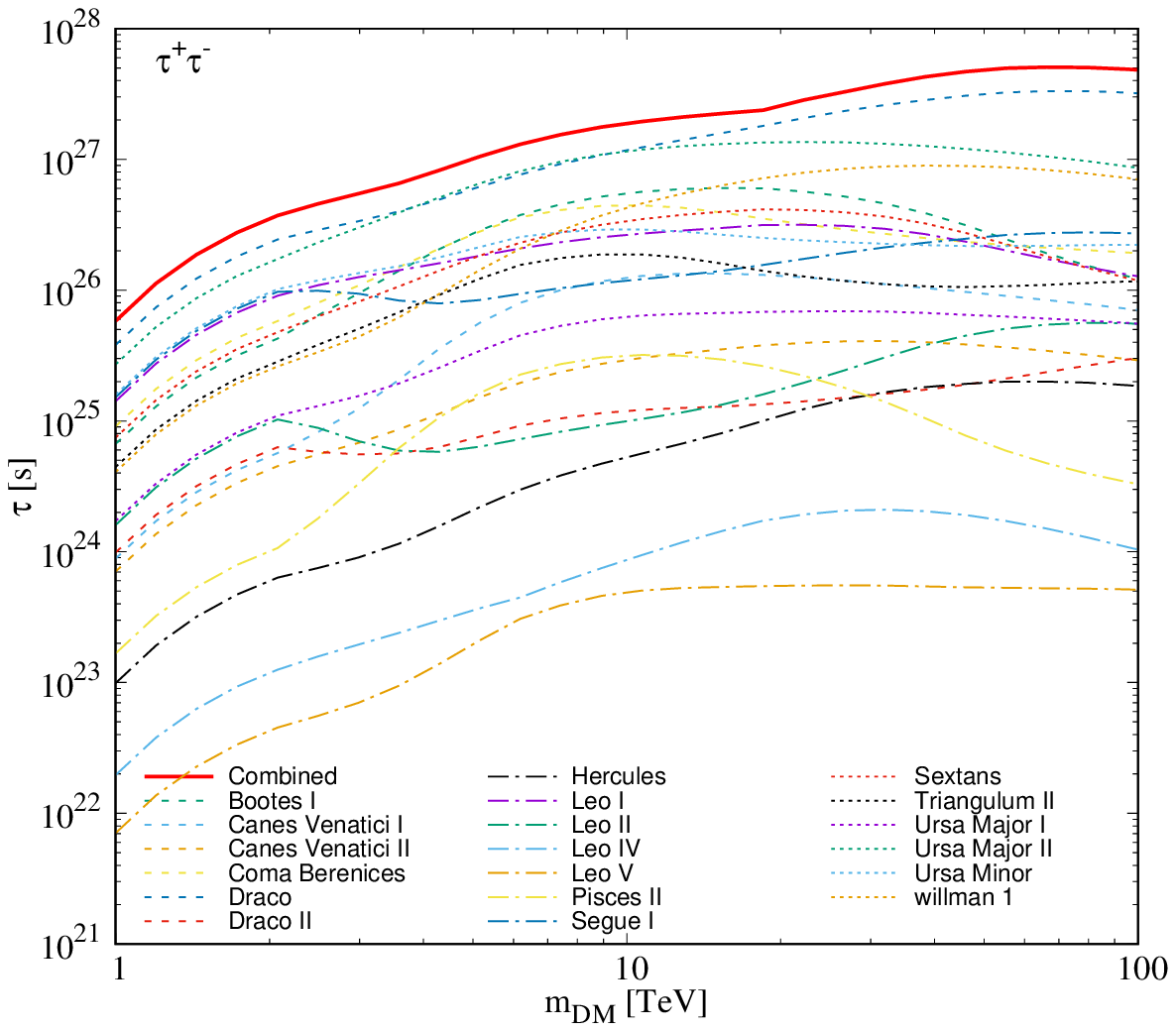}}
{\includegraphics[width=0.45\textwidth]{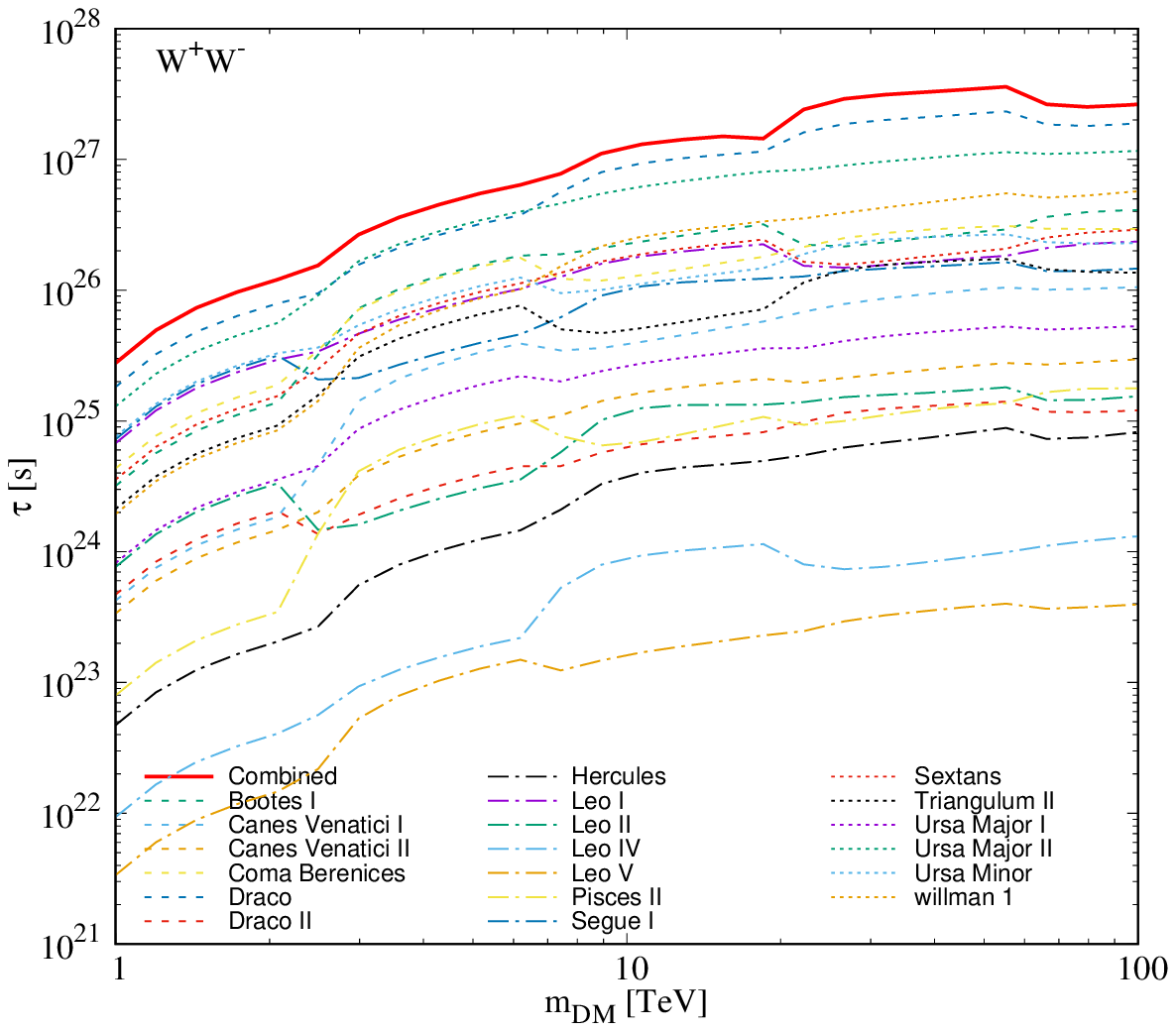}}
\end{center}
\caption{Projected one-year LHAASO sensitivities with respect to DM lifetime $\tau$ at $95\%$ confidence level for individual dSphs in one mimic observation.
Five decay channels $ b\bar{b}$, $t\bar{t}$, $\mu^{+}\mu^{-}$, $\tau^+\tau^-$, and $\rm W^+W^-$ are considered.
The solid red line represents the combined sensitivity with all dSphs.}
\label{fig:Lhaaso-separate}
\end{figure*}

Fig. \ref{fig:Lhaaso-separate} shows the LHAASO sensitivities with respect to the DM lifetime for the individual dSph in a randomly selected mimic observation.
Here, we consider five typical DM decay channels, including $b\bar{b}$, $t\bar{t}$, $\mu^{+}\mu^{-}$, $\tau^+\tau^-$, and $\rm W^+W^-$.
In this figure, the LHAASO sensitivities from a combined analysis with all selected dSphs are likewise shown, using a joint likelihood $\mathcal{L}^{\rm tot}=\prod_{j}\mathcal{L}_{j}$.
We see that the combined sensitivity is dominated by two dSphs, {Draco} and {Ursa Major II}, with large $D$ factors and favorable locations inside the LHAASO FOV.
Notably, {Triangulum II} and {Willman 1} have almost the largest $D$ factors among the 19 selected dSphs, and t{Triangulum II} is located near the center of LHAASO FOV.
However, their contributions do not significantly affect the combined sensitivity.
The reason is that the statistical uncertainties of their $D$ factors are considerably large, owing to the lack of data from the dSph kinematic observations.
The sensitivity to the gamma-ray signal from the dSph would be significantly decreased by including the uncertainty of the $D$ factor in the likelihood.
Nevertheless, although some dSphs have relatively large $D$ factors with small uncertainties, they are close to the edge of LHAASO FOV.
Consequently, the signals from these dSphs are difficult to detect by LHAASO.

\begin{figure*}
\begin{center}
	{\includegraphics[width=0.45\textwidth]{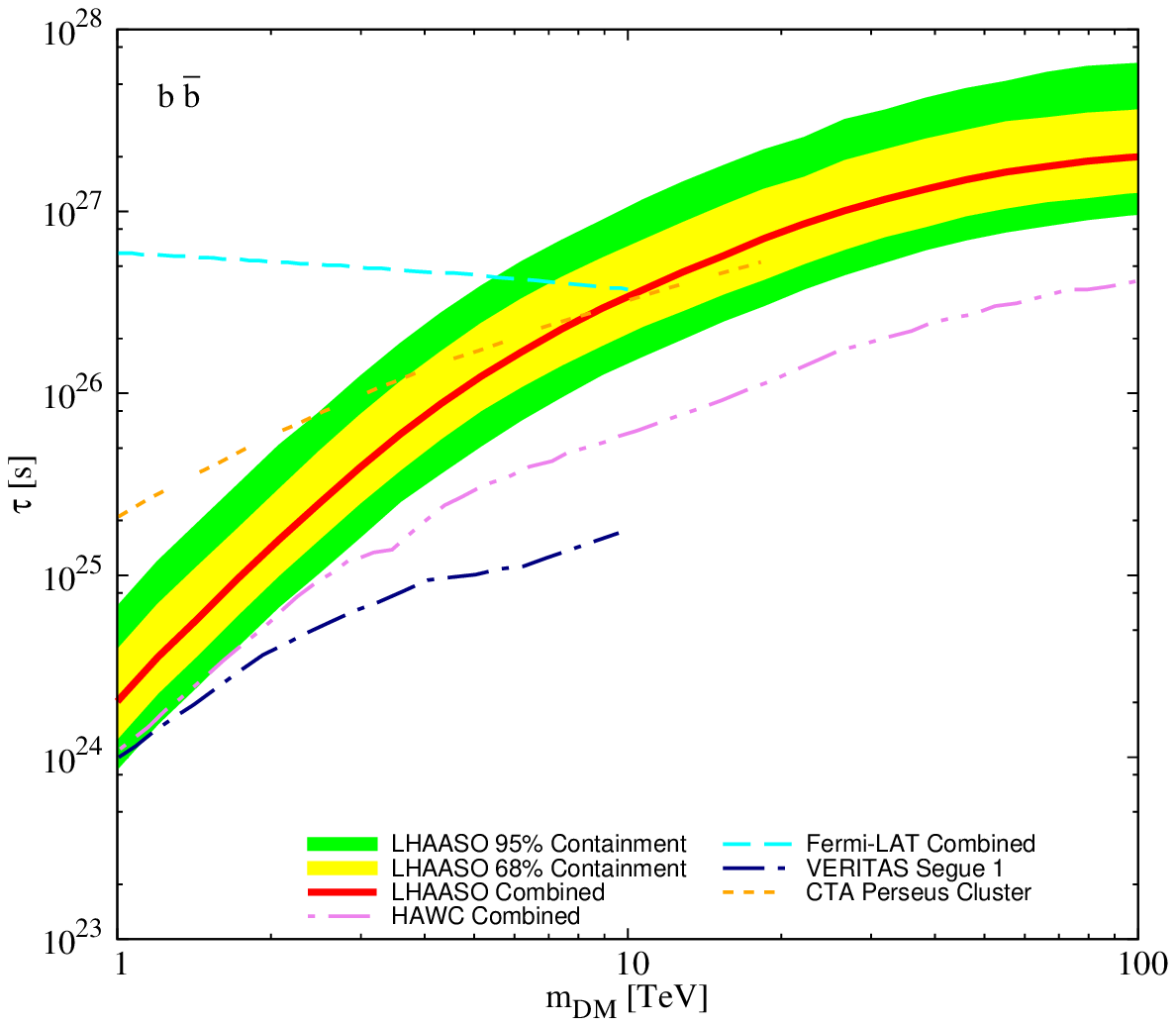}}
	{\includegraphics[width=0.45\textwidth]{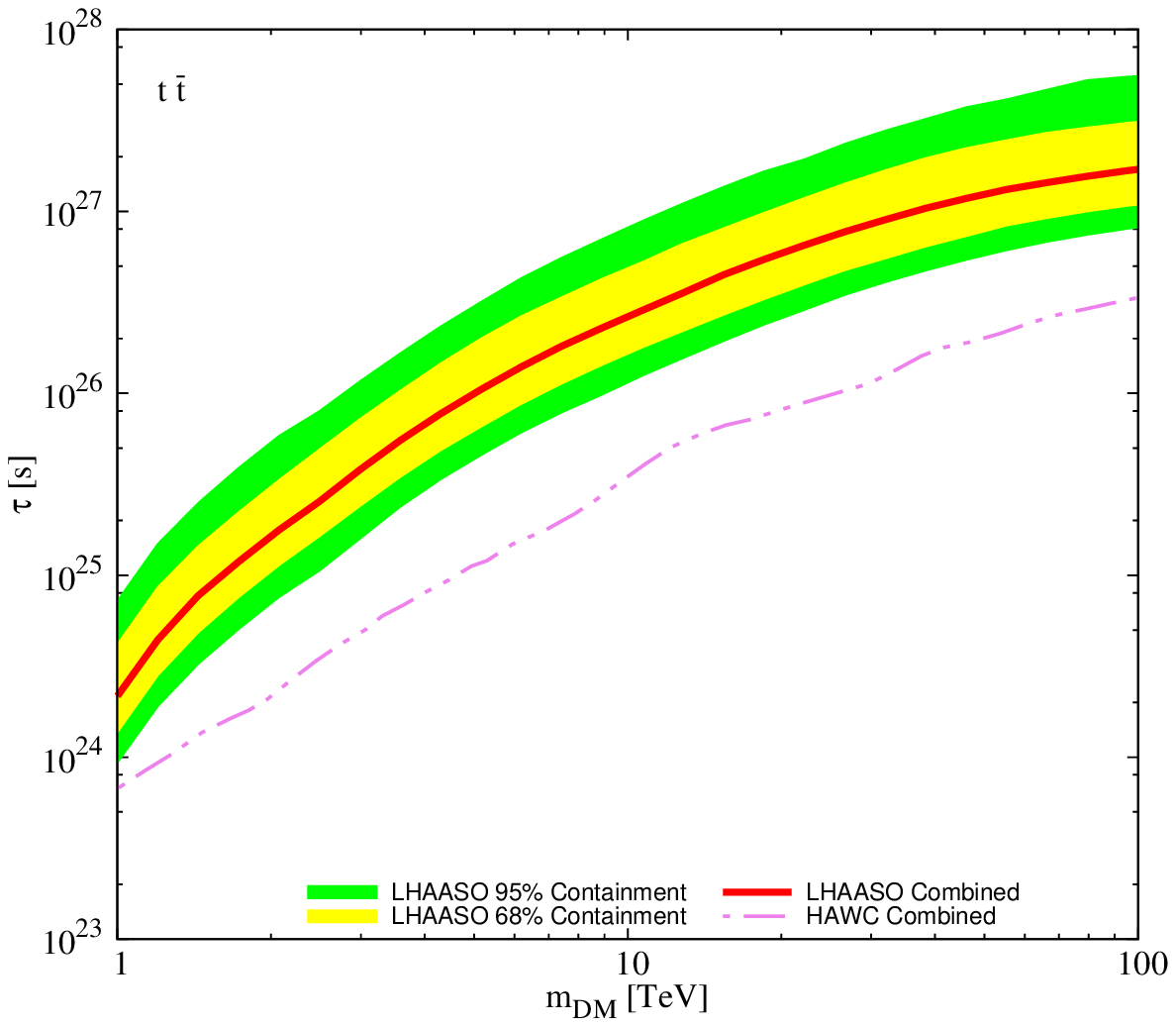}}
	{\includegraphics[width=0.45\textwidth]{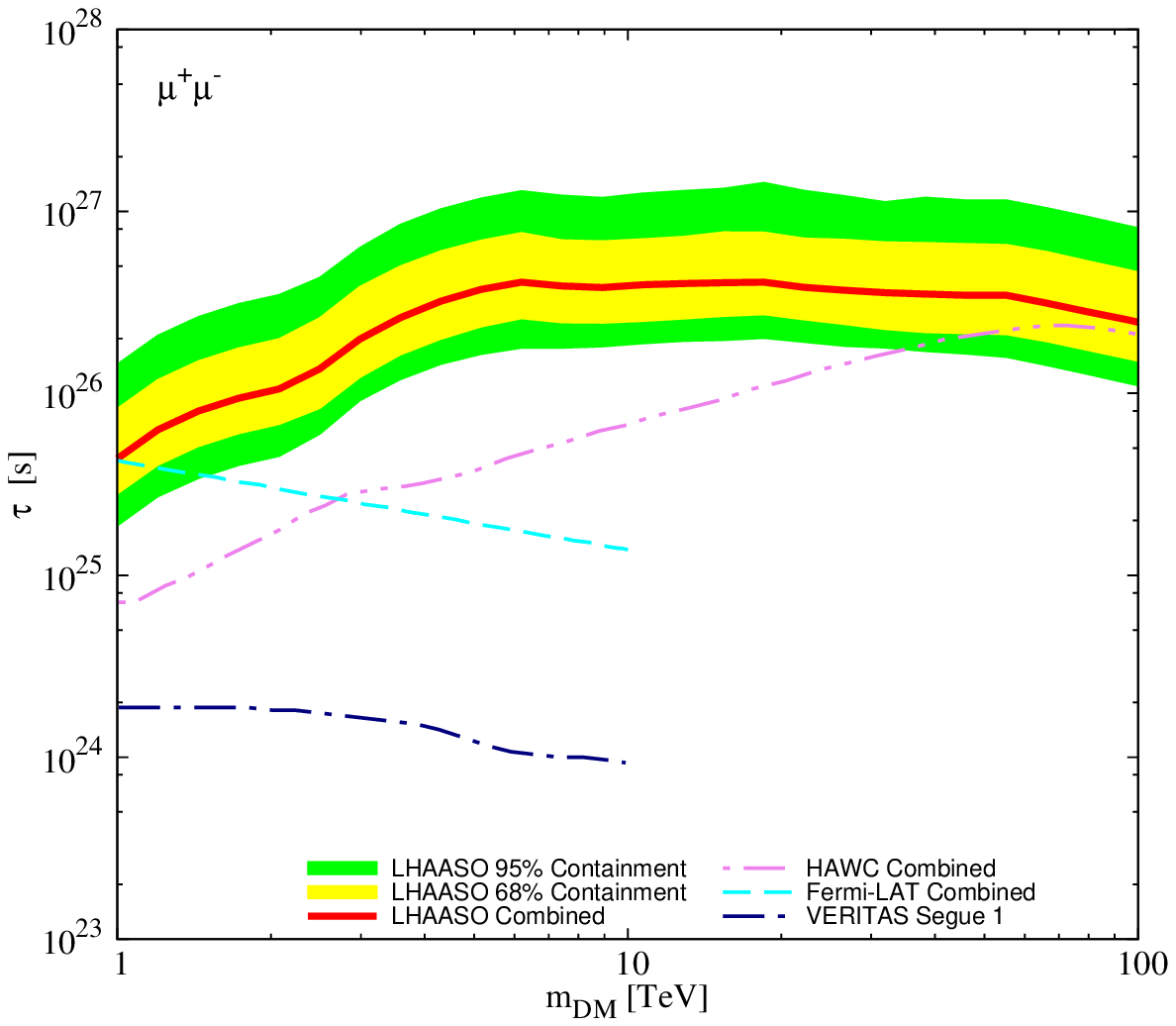}}
	{\includegraphics[width=0.45\textwidth]{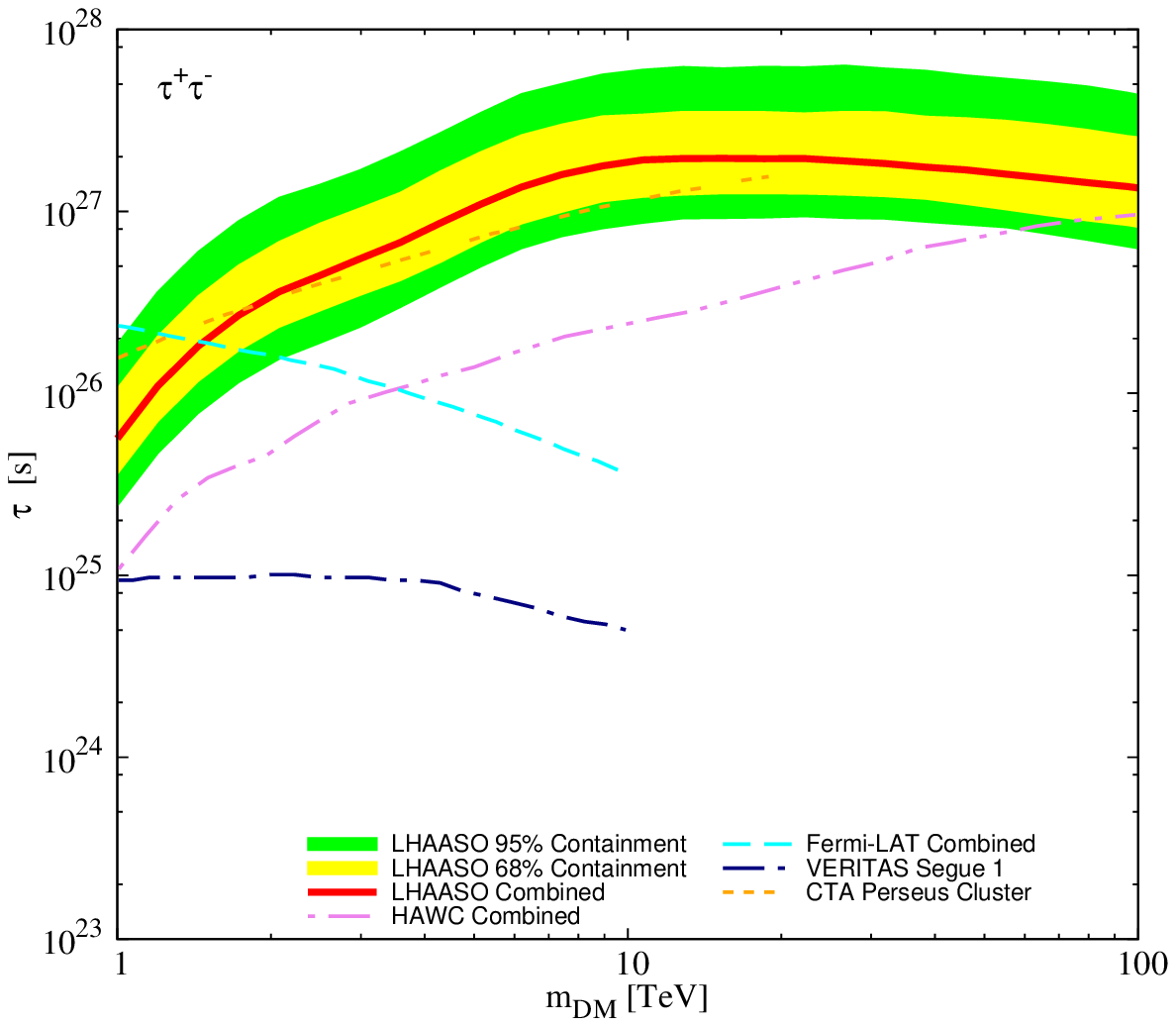}}
	{\includegraphics[width=0.45\textwidth]{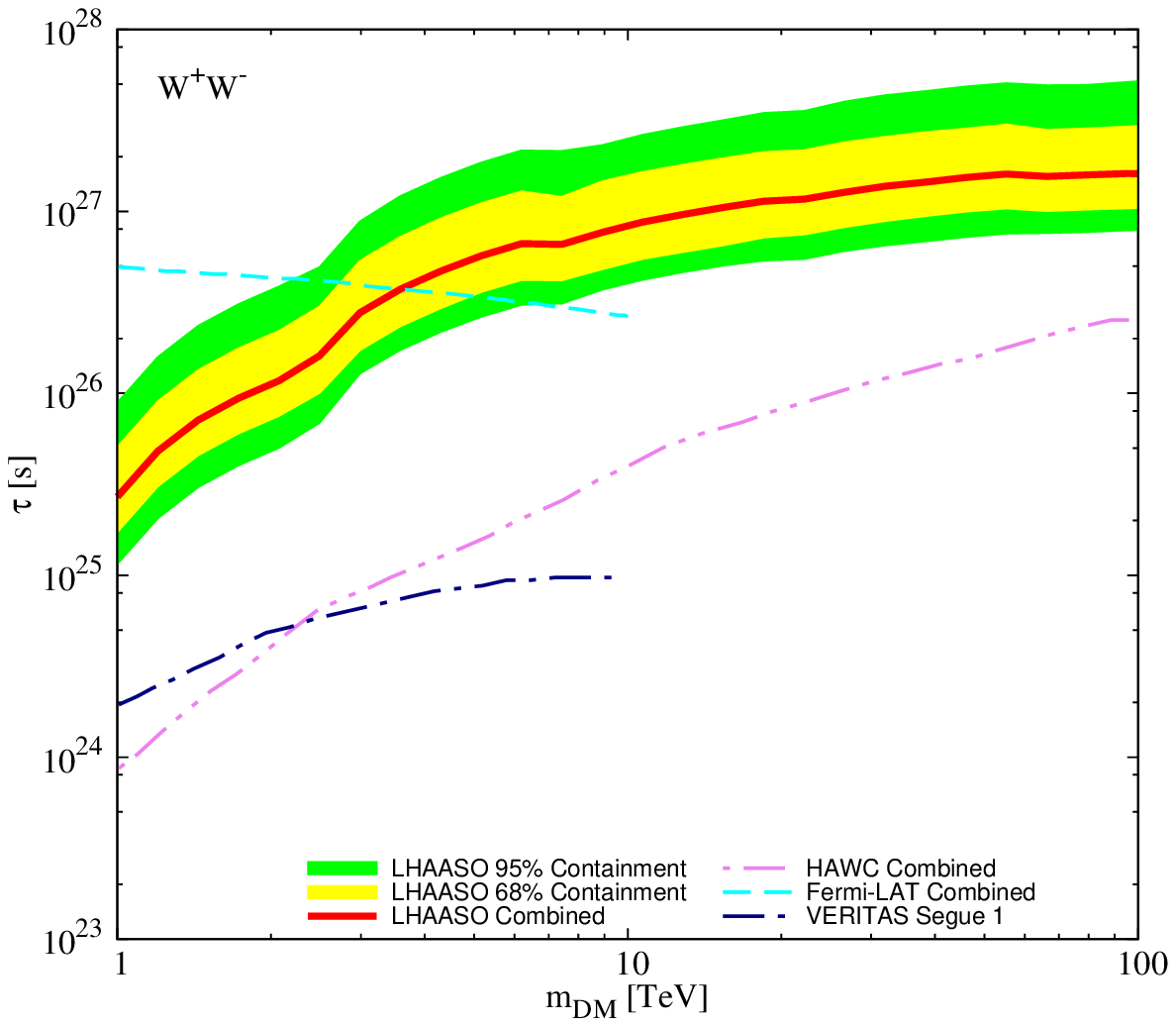}}
\end{center}
\caption{Expected one-year combined sensitivities of LHAASO for five decay channels including $b\bar{b}$, $ t\bar{t}$, $\mu^{+}\mu^{-}$, $\tau^+\tau^-$, and $\rm W^+W^-$.
Both median values (red solid lines) and related two-sided $68\%$ (yellow) and $95\%$ (green) containment bands of the sensitivities are shown.
The limits from some other experiments are depicted for comparison, including the HAWC combined dSph limit \cite{Albert:2017vtb}, Fermi-LAT combined dSph limit \cite{Baring:2015sza}, VERITAS Segue 1 limit \cite{Aliu:2012ga}, {and CTA Perseus Cluster prospective limit \cite{Acharya:2017ttl,Morselli:2017ree}}.
}
\label{fig:Lhaaso-combined}
\end{figure*}

Because there are statistic fluctuations in each mimic observation, we perform 500 mimic observations under the null hypothesis to include this uncertainty in the final results.
We show the median combined sensitivities and the related two-sided 68\% and 95\% containment bands for the five decay channels in Fig.~\ref{fig:Lhaaso-combined}.
For comparison, the lower limits on the DM lifetime from three gamma-ray observations, including the HAWC combined dSphs limit \cite{Albert:2017vtb}, Fermi-LAT combined dSphs limit \cite{Baring:2015sza}, VERITAS Segue 1 limit \cite{Aliu:2012ga}, {and the prospective limits of CTA Perseus Cluser observation \cite{Acharya:2017ttl,Morselli:2017ree}}, are also shown.

Fig.~\ref{fig:Lhaaso-combined} shows that the LHASSO sensitivities with respect to DM decay lifetime are better than the current experimental limits for the DM masses $m_\chi$ approximately larger than $10$ TeV and reach $\mathcal{O}(10^{27} \,\rm s)$ for almost all channels at the mass scale of $1 - 100$ TeV.
For hadronic channels, such as the $b\bar{b}$ channel, the initial photon spectra are soft.
Thus, the Fermi-LAT dSph observations place the most stringent limits for DM masses approximately up to $10$ TeV, due to their considerably good sensitivities to the low-energy gamma rays.
For DM masses below 10 TeV, as we impose a cut on the observed photon energies $E>0.7$ TeV, which is consistent with the public LHASSO simulation results, the expected LHAASO sensitivities in this energy region are poor compared with the current limits.
Nevertheless, for the masses above 10 TeV, LHAASO could become more sensitive for these channels through the excellent observation of VHE photons.
Moreover, for the $W^{+}W^{-}$ channel, LHAASO behaves as the most sensitive experiment, in comparison with the limits set by other experiments, at masses above $\simeq3$ TeV.
With regard to the $\mu^{+}\mu^{-}$ and $\tau^{+}\tau^{-}$ channels, LHAASO exhibits excellent sensitivities for almost all DM masses above $1$ TeV.

We furthermore find that, although including the statistical uncertainties of $D$ factors in our analysis would decrease the LHAASO sensitivity, the expected combined sensitivities are nevertheless better than the limits set by HAWC by a factor of $8-10$.
As shown in Ref. \citep{He:2019dya}, however, the LHAASO sensitivities to the DM annihilation cross section are only stronger than the HAWC limits by a factor of $2-5$.
Evidently, the improvement for DM decay is even more significant than the case of DM annihilation.

We discuss the potential reasons for this factor.
In Fig. \ref{fig:Lhaaso-separate}, we clearly see that the combined sensitivity is in particular dominated by high-latitude sources, such as {Draco} and {Ursa Major II}.
For these sources, the LHAASO sensitivities would be significantly stronger than HAWC by 1$-$2 orders of magnitude.
This is because LHAASO is located at a higher latitude of approximately $29^{\circ}$ in comparison with HAWC, which is located at the latitude of approximately 19$^{\circ}$.
The dSphs at high latitudes, such as {Draco} and {Ursa Major II}, are located near the edge of the HAWC FOV, such that HAWC is insensitive to them.
However, these sources could still contribute significant signals in the LHAASO FOV.
In addition, LHAASO has a larger effective area in comparison with HAWC.
Therefore, the LHAASO sensitivities to the DM decay lifetime are conceivable to be higher than the constraints from HAWC by a factor of $8-10$.
%Furthermore, as the DM decay is related to $\int\rho$ compared with $\int\rho^2$ in the case of DM annihilation, the $D$ factors of the sources would exert a different hierarchy of influence on the combined sensitivity than the $J$-factors for annihilation.

\begin{table*}[!h]
\caption{Number of the expected background events for each dSph.}
\footnotesize\centering
\label{back}

\begin{tabular*}{170mm}{@{\extracolsep{\fill}}ccccccccccccccc}\toprule
          	&&	
        0.7 TeV$-$2.1 TeV&&
		2.1 TeV$-$6.3 TeV&&
		6.3 TeV$-$18.9 TeV&&
        18.9 TeV$-$56.7 TeV\\\hline

		$\rm Bo\ddot{o}tes ~I$&&
         185068&&
		5407&&
		1747&&
		765\\

		Canes Venatici $\rm I$&&
		209117&&
		6163&&
		2233&&
		977\\
		
		Canes Venatici~$\rm II$&&
		209929&&
		6133&&
		1678&&
        628\\
		
		Coma~Berenices&&
		197879&&
		5781&&
		1851&&
        592
		\\
		
		Draco&&
		231368&&
		6823&&
		2473&&
        1083
		\\

		Draco II&&
		236163&&
		6965&&
		2526&&
        1107
		\\
		
        Hercules&&
		182507&&
		5333&&
		1459&&
        546
		\\

        Leo I&&
		181748&&
		5311&&
		1592&&
        689\\

        Leo II&&
		195646&&
		5716&&
	    1564&&
        585\\

        Leo IV&&
		159002&&
		4648&&
		1272&&
        476\\

        Leo V&&
		164471&&
		4807&&
		1316&&
        493\\

        Pisces II&&
		171309&&
		5051&&
		1831&&
        802\\

        Segue 1&&
		187371&&
		5475&&
		1498&&
        560\\

        Sextans&&
		156774&&
		4623&&
		1676&&
        734\\

        Triangulum II&&
		211880&&
		6245&&
		2262&&
        990\\

        Ursa Major I&&
		226549&&
		6621&&
		1878&&
        784\\

        Ursa Major II&&
		235186&&
		6936&&
		2515&&
        1102\\

        Ursa Minor&&
		237888&&
		7017&&
		2545&&
        1115\\

        Willman 1&&
		225815&&
		6658&&
		2413&&
        1057\\
    \bottomrule
    %\hline\hline
	\end{tabular*}
    \end{table*}

\section{Conclusion}\label{sec:conclu}
We study the sensitivities of the LHAASO dSph observations to the lifetime of decaying DM particles for five final decay states.
Both individual and combined limits for 19 dSphs incorporating the statistical uncertainties of $D$ factors are investigated.
Our results show that the contributions of two sources, dSphs {Draco} and the {Ursa Major II}, significantly affect the combined sensitivity.
In comparison with the current limits from Fermi-LAT, HAWC, and VERITAS, we find that the LHAASO sensitivities are better for the DM masses larger than approximately 3 TeV and 10 TeV for the $\rm W^{+}W^{-}$ and $b\bar{b}$ channels, respectively.
Furthermore, LHAASO is sensitive in a wide DM mass range  from 1 to 100 TeV.
Therefore, we conclude that the LHAASO dSph gamma-ray observation would be a compelling and promising approach for probing the properties of heavy decaying DM particles above $\mathcal{O}(\rm TeV)$.

\appendix
\section*{Appendix A: Expected event counts at LHAASO} \label{section:method}
We perform a series of mimic observations to derive the expected LHAASO sensitivities with respect to the DM decaying signals.
First, we estimate the expected background counts $B$ induced by cosmic ray nuclei.
Second, we perform a Gaussian sampling around $B$ to obtain the observational event counts $N$ in each mimic observation.

The energy resolution of WCDA varies from 30\% to 100\% with the decreasing energy.
We adopt sufficiently wide energy bins with $E_{\rm max}/E_{\rm min}=3$, such that the energy smearing effect can be ignored in our analysis.
The background count $B$ in one energy bin is estimated by
\begin{equation}\label{func:bkg}\small
B=\zeta_{cr}\int^{E_{\rm max}}_{E_{\rm min}}\int_{\Delta\Omega}\int^{T}_{0}\Phi_{p}(E)\cdot A_{\rm eff}^{p}(E,\theta_{\rm zen}(t))\cdot\varepsilon_{p}(E)dtd\Omega dE,
\end{equation}
where $\Phi_{p}(E)$ is the primary proton flux in cosmic rays, which is taken to be a single power-law from the fitting to the results of ATIC \cite{Panov:2011ak}, CREAM \cite{Yoon:2011aa}, and RUNJOB \cite{Derbina:2005ta}.
The total observational time of LHAASO for this analysis is taken to be one year.
The number of the event counts is estimated within a cone of $\Delta\Omega=2\pi\times[1-\cos(\max\{ \alpha_{\rm int}, \theta_c \})]$, where $\theta_c$ denotes the energy dependent angular resolution of LHAASO, varying from $2^\circ$ to $0.1^\circ$ with the increased energy of the gamma-ray \citep{Bai:2019khm}.
Here, we also introduce a scale factor $\zeta_{cr}=1.1$ to include the contributions of other heavy nuclei in the primary cosmic rays.

The expected signal event count $S$ in one energy bin is calculated by
\begin{equation}\label{func:sig}
S=\epsilon_{\Delta\Omega}\int^{E_{\rm max}}_{E_{\rm min}}\int_0^T\Phi_{\gamma}(E)\cdot A_{\rm eff}^{\gamma}(E,\theta_{\rm zen}(t))\cdot\varepsilon_{\gamma}(E)dtdE,
\end{equation}
where $\epsilon_{\Delta\Omega}=0.68$ denotes the fraction of observed photons within the experimental angular resolution.

The effective area of LHAASO $A_{\rm eff}^{p}$ is a function of the energy and zenith angle.
Here, we take $A_{\rm eff}^{p}$ from the LHAASO science white paper \cite{Bai:2019khm}.
Notably, the zenith angle $\theta_{\rm zen}$ is also a function of the observation time $t$.
DSphs at different declinations are expected to have different $\theta_{\rm zen}(t)$ functions, leading to different visibilities.
To reflect the visibility, we show the effective time ratio $r_{\rm eff}$ in Table~\ref{Table:dsphs}.
This factor denotes the fraction of effective observation time during which the corresponding zenith angle $\theta_{\rm zen}$ of the dSph is smaller than $60^\circ$.

In the above formulae, $\varepsilon$ denotes the survival ratio of the particle after selection in the experimental analysis, which reflects the efficiency for the gamma-proton discrimination.
A detailed analysis is provided in Ref. \cite{Zha:2017vcs} for the working efficiencies of WCDA.
This analysis shows that for the energies above $0.6\,\TeV$, the survival rate of the proton $\varepsilon_{p}$ can be suppressed in a range from 0.04\% to 0.11\%, while the survival rate of the gamma $\varepsilon_{\gamma}$ is approximately $50\%$.
In this study, we adopt a more conservative gamma-proton discrimination as $\varepsilon_{p}\simeq 0.28\%$ with $\varepsilon_{\gamma} \simeq 40.13\%$.

{Taking account of all these factors, we list the expected background count $B$ for each dSph in Table~\ref{back}.}

\vspace{3mm}
\providecommand{\href}[2]{#2}\begingroup\raggedright\endgroup

\end{multicols}

%\clearpage

%\end{CJK*}
\end{document}